\documentclass[3p,times]{elsarticle}
\usepackage{natbib}
\usepackage[figuresright]{rotating}

\begin{document}

\begin{frontmatter}

\title{On the Hydrogen Atom Beyond the Born-Oppenheimer Approximation}

\author[bg]{Jean~Michel~Sellier$^*$}
\author[bg]{K.G.~Kapanova}
\address[bg]{IICT, Bulgarian Academy of Sciences, Acad. G.~Bonchev str. 25A, 1113 Sofia, Bulgaria\\$^*$\texttt{jeanmichel.sellier@parallel.bas.bg,\\jeanmichel.sellier@gmail.com}}

\begin{abstract}
Recently a new formulation of quantum mechanics has been suggested which is based on the concept of {\sl{signed particles}},
that is, classical objects provided with a position, a momentum and a sign simultaneously.
In this paper, we comment on the plausibility of simulating atomic systems beyond the Born-Oppenheimer approximation
by means of the signed particle formulation of quantum mechanics. First, in order to show the new perspective offered
by this new formalism, we provide an example studying quantum tunnelling through a simple Gaussian barrier in terms
of the signed particle formulation.
Then, we perform a direct simulation of the hydrogen atom as a full quantum two-body system, showing that the
formalism can be a very promising tool for {\sl{first-principle-only}} quantum chemistry.\end{abstract}


\end{frontmatter}

\section{Introduction}

The scientific exploration of the hydrogen atom (consisting of one electron and one proton) has 
had a profound effect in the development of the quantum theory.  In 1913, N.~Bohr provides
the first quantum model in history, theoretically describing the observed quantum energetic
jumps experienced by electrons in an atom. The theory was based on the Plank's concept of quanta,
and therefore still constrained to the orthodoxy of classical physics. The Schr\"{o}dinger equation
was later applied to a system, providing a way to discover exactly the electronic energies and
orbitals of hydrogen.

A key part of the advances in the field is the introduction of one important approximation,
based on the assumption that the motion of nucleus and electrons can be separated. Suggested
in $1927$ by M.~Born and J.R.~Oppenheimer, the  physical basis 
for approximation is grounded on the fact that the mass of the proton
is considerably higher than that of an electron (about $1836$ times the electron mass), which makes
the nuclei to move much slower, allowing for its kinetic energy to be neglected \cite{BO}.

Despite its incredible success in both quantum physics and chemistry, it is known that the Born-Oppenheimer
approximation disregards some relevant physical processes. For instance, the effects due to vibronic
coupling, responsible for describing the influence of electronic vibrations over the nucleus (and vice-versa),
are entirely neglected, and as a consequence limit the study of non-adiabatic processes.
Moreover, due to the very different time scales involved in electronic and proton transport,
the treatment of an atom as a full quantum many-body object (electrons {\sl{and}} protons
all together) remains a challenging computational problem \cite{MD}, \cite{NQC}.

Recently, a new formulation of quantum mechanics based on the concept of signed particles has
been introduced \cite{SPF}. The underlying principle of the formulation utilizes ensembles of
classical signed field-less particles in the simulation of time-dependent quantum system.
The formalism consist of a set of three postulates, which describe the evolution of a
quantum system. Specifically, these rules dictate how an ensemble of particles evolves,
one particle at a time (which allows for trivial parallelization since in this context
particles are completely independent from each other). This approach is completely equivalent
to other (more standard) formulations of quantum mechanics such as, for example, the one
suggested by E.~Schr\"{o}dinger (in other words, they provide {\sl{exactly the same predictions}}).
Importantly, the formulation has been extended to time-dependent, quantum many-body problems
showing already to be extremely promising and useful, e.g. in the case of indistinguishable
Fermions \cite{JCP-01}, \cite{JCP-02}, \cite{JCP-03}. 

The purpose of this work is to apply for the first time the signed Particle formalism to the hydrogen atom
considered as a time-dependent two-body problem, therefore going beyond the Born-Oppenheimer approximation.
To prove our point, we stress on the fact that {\sl{both}} the electron and proton are considered as quantum
objects and the system is treated as a {\sl{whole}} described by the Signed particle formalism.

The paper is organized as follows. In the next section, we introduce the formulation and we present the set
of postulates which completely defines the new theory. In order to highlight the new perspective offered
by this novel theory, we discuss the formalism in the context of the 
well-known quantum tunnelling effect. Finally, we show preliminary one-dimensional results from a simulation of a 
hydrogen atom, showing that quantum physics and chemistry beyond the Born-Oppenheimer approximation is
at our reach at a level of details never attained before.

\section{The signed particle formulation}

This section introduces the signed particle formulation of quantum mechanics recently suggested in \cite{SPF}.
It consists of a set of three postulates which are reported below. These postulates, or rules, can be seen as physical laws defining
the dynamics of (classical) {\sl{signed}} particles in the presence of a given potential. In this context,
quantum systems are now described by means of {\sl{ensembles}} of signed particles.
For the sake of simplicity, we limit the discussion to the case of a one-dimensional real space, although the generalization
to higher-dimensional space is possible and relatively easy to obtain.

\bigskip

{\sl{{\bf{Postulate I.}} Physical systems can be described by means of (virtual) Newtonian particles, i.e. provided with a
position ${\bf{x}}$ and a momentum ${\bf{p}}$ simultaneously, which carry a sign which can be positive or negative.}}

\bigskip

{\sl{{\bf{Postulate II.}} A signed particle, evolving in a potential $V=V \left( {\bf{x}} \right)$, behaves as a
field-less classical point-particle which, during the time interval $dt$, creates a new pair of signed particles
with a probability $\gamma \left( {\bf{x}}(t) \right) dt$ where
\begin{equation}
 \gamma\left( {\bf{x}} \right) = \int_{-\infty}^{+\infty} \mathcal{D}{\bf{p}}' V_W^+ \left( {\bf{x}}; {\bf{p}}' \right)
\equiv \lim_{\Delta {\bf{p}}' \rightarrow 0^+} \sum_{{\bf{M}} = -\infty}^{+\infty} V_W^+ \left( {\bf{x}}; {\bf{M}} \Delta {\bf{p}}' \right),
\label{momentum_integral}
\end{equation}
and $V_W^+ \left( {\bf{x}}; {\bf{p}} \right)$ is the positive part of the quantity
\begin{equation}
 V_W \left( {\bf{x}}; {\bf{p}} \right) = \frac{i}{\pi^d \hbar^{d+1}} \int_{-\infty}^{+\infty} d{\bf{x}}' e^{-\frac{2i}{\hbar} {\bf{x}}' \cdot {\bf{p}}} \left[ V \left( {\bf{x}}+{\bf{x}}'\right) - V \left( {\bf{x}}-{\bf{x}}'\right)  \right],
\label{wigner-kernel}
\end{equation}
known as the Wigner kernel (in a $d$-dimensional space) \cite{Wigner}.
If, at the moment of creation, the parent particle has sign $s$,
position ${\bf{x}}$ and momentum ${\bf{p}}$,
the new particles are both located in ${\bf{x}}$, have signs $+s$ and $-s$,
and momenta ${\bf{p}}+{\bf{p}}'$ and ${\bf{p}}-{\bf{p}}'$ respectively,
with ${\bf{p}}'$ chosen randomly according to the (normalized)
probability $\frac{V_W^+ \left( {\bf{x}}; {\bf{p}} \right)}{\gamma({\bf{x}})}$.}}

\bigskip

{\sl{{\bf{Postulate III.}} Two particles with opposite sign and same phase-space coordinates $\left( {\bf{x}}, {\bf{p}}\right)$ annihilate.}}

\bigskip

The obtained physical picture by this novel approach is rather peculiar compared to the standard
quantum mechanics (albeit there is certain resemblance to the Feynman approach). In point of fact,
a quantum system is forthwith described by ensembles of field-less, classical (i.e. provided with a
position and a momentum simultaneously) particles. In the formalism those point-like objects
carry a sign which can be either positive or negative. Their interaction with the potential $V=V(x)$
is realized through particle creation events only. Accordingly, one obtains
the macroscopic properties of a quantum system by averaging a macroscopic variable
$A=A\left( x; p\right)$ over the ensemble of particles, taking into account their sign, i.e.
\begin{equation}
 <A> = \frac{1}{N} \sum_{i=1}^N s_i A\left( x_i; p_i \right),
\label{macroscopic-value}
\end{equation}
where $s_i$, $x_i$ and $p_i$ are represented by the sign, the position and the momentum, respectively, of the $i$-th particle
(just as in classical statistical mechanics, but now taking the sign of the particles into account).

The postulates are derived from a physical interpretation of the time-dependent Wigner equation which is
rewritten in terms of a Fredholm equation of second kind \cite{SPF}, \cite{PhysRep}. It is possible to prove that
this set of rules is sufficient to reconstruct the time-dependent evolution of any quantum system \cite{SPF}.
In a similar way, one can extend this set of postulates to the case of many-body systems (starting, this time, from the
{\sl{many-body}} time-dependent Wigner equation)\footnotemark.

\footnotetext[1]{The interested reader is encouraged to download and study the source code implementing this algorithm from \cite{nano-archimedes}.}

\begin{figure}[h!]
\centering
\begin{minipage}{0.99\textwidth}
\begin{tabular}{c}
\includegraphics[width=0.45\textwidth]{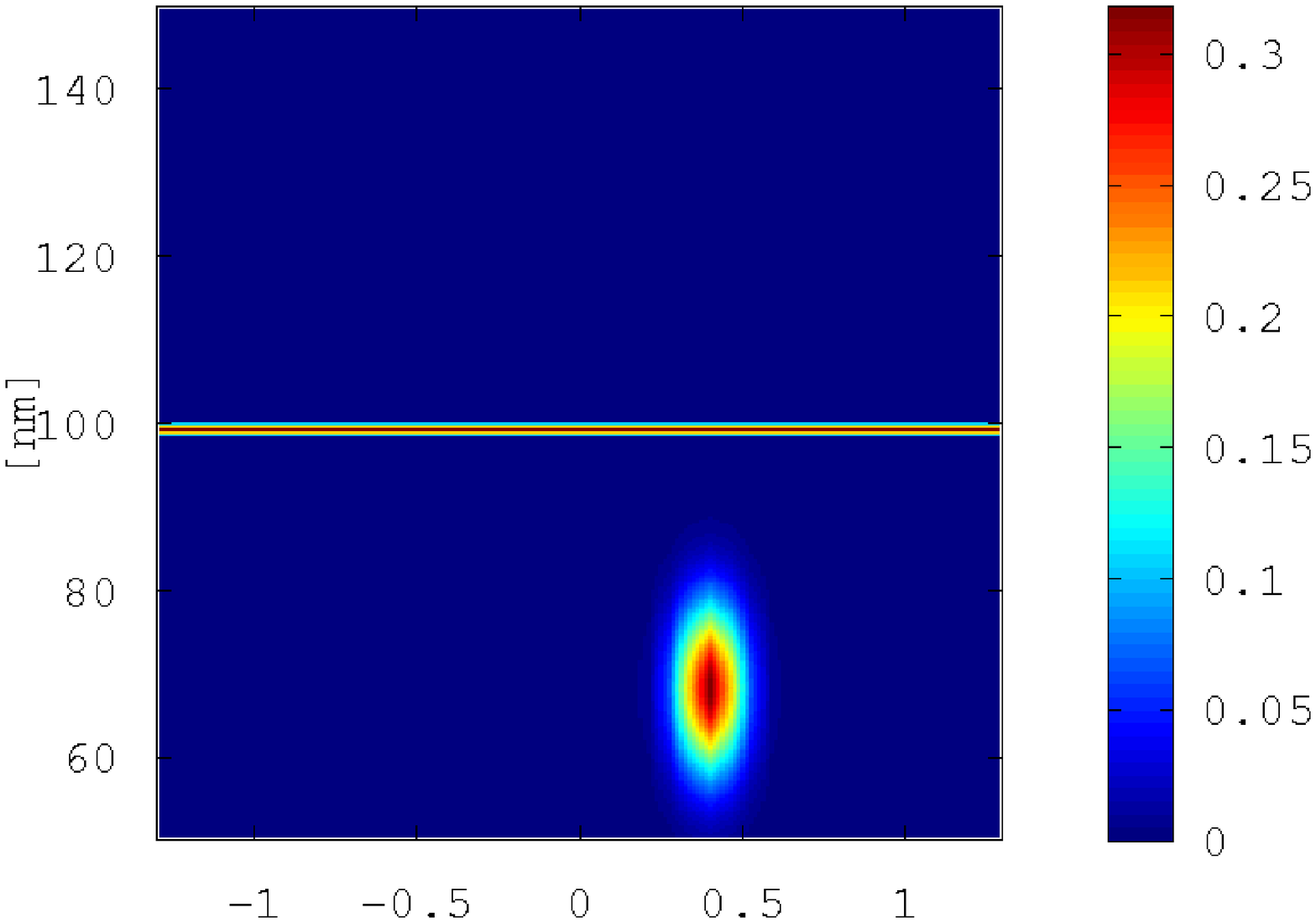}
\includegraphics[width=0.45\textwidth]{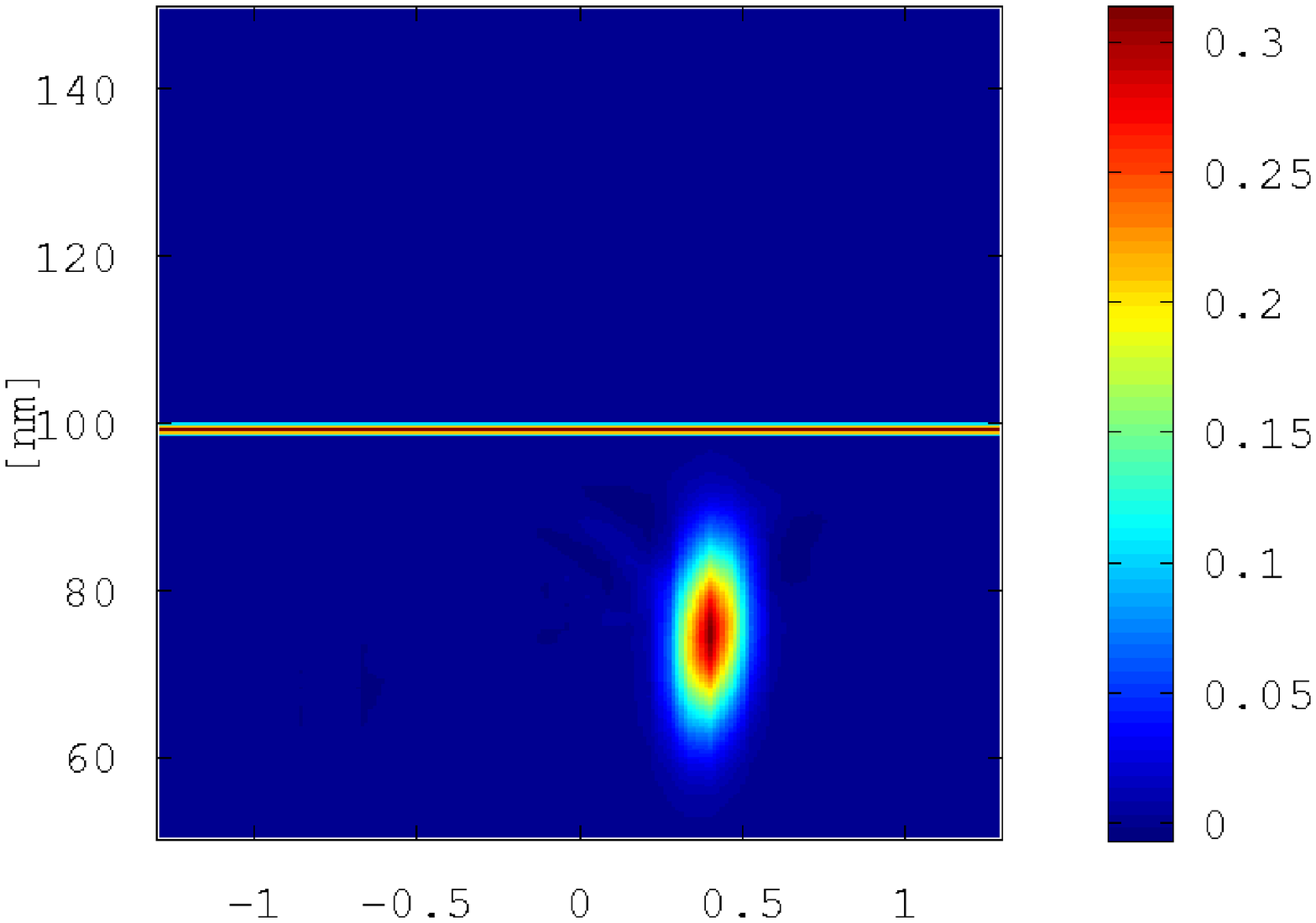}
\\
\includegraphics[width=0.45\textwidth]{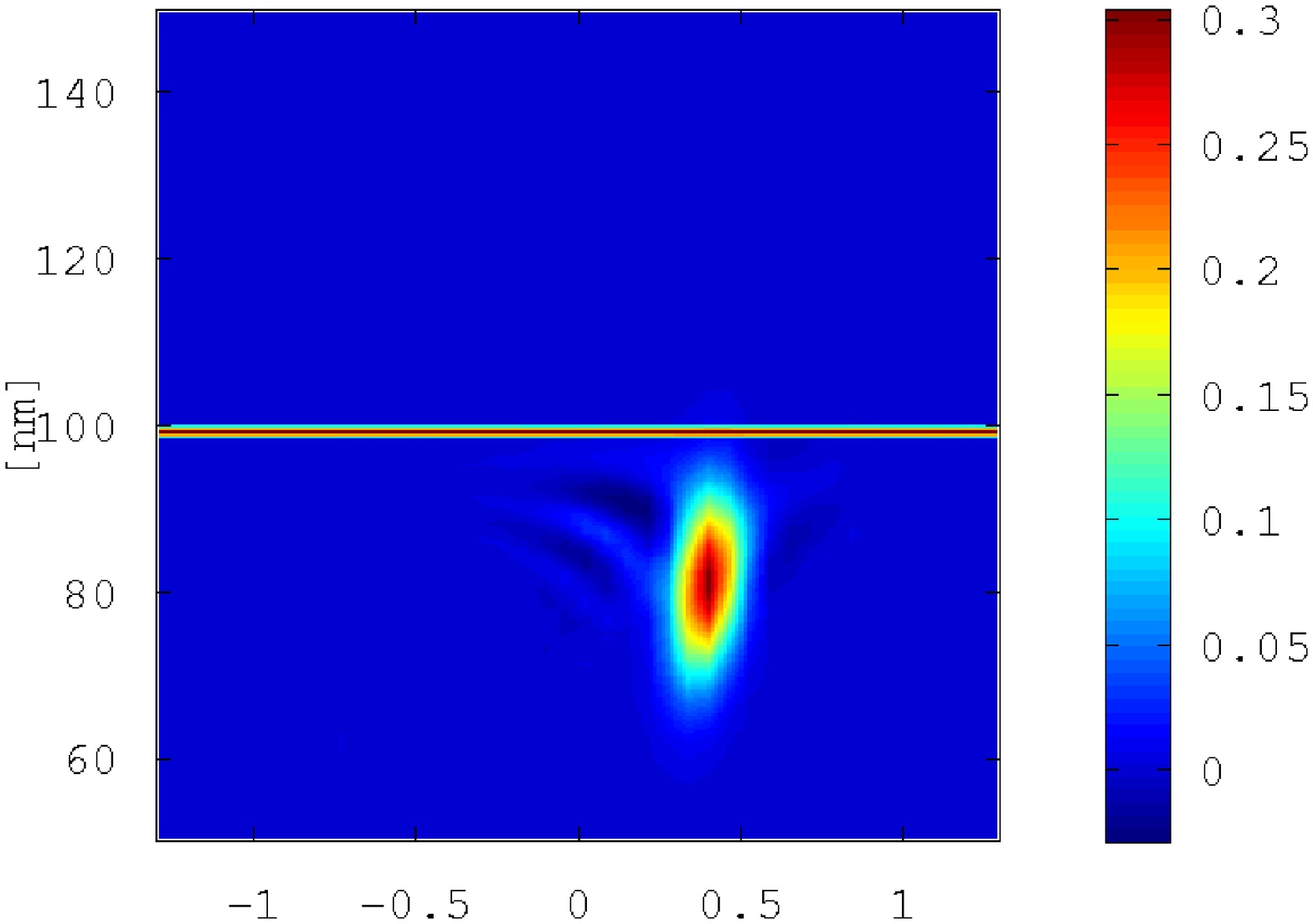}
\includegraphics[width=0.45\textwidth]{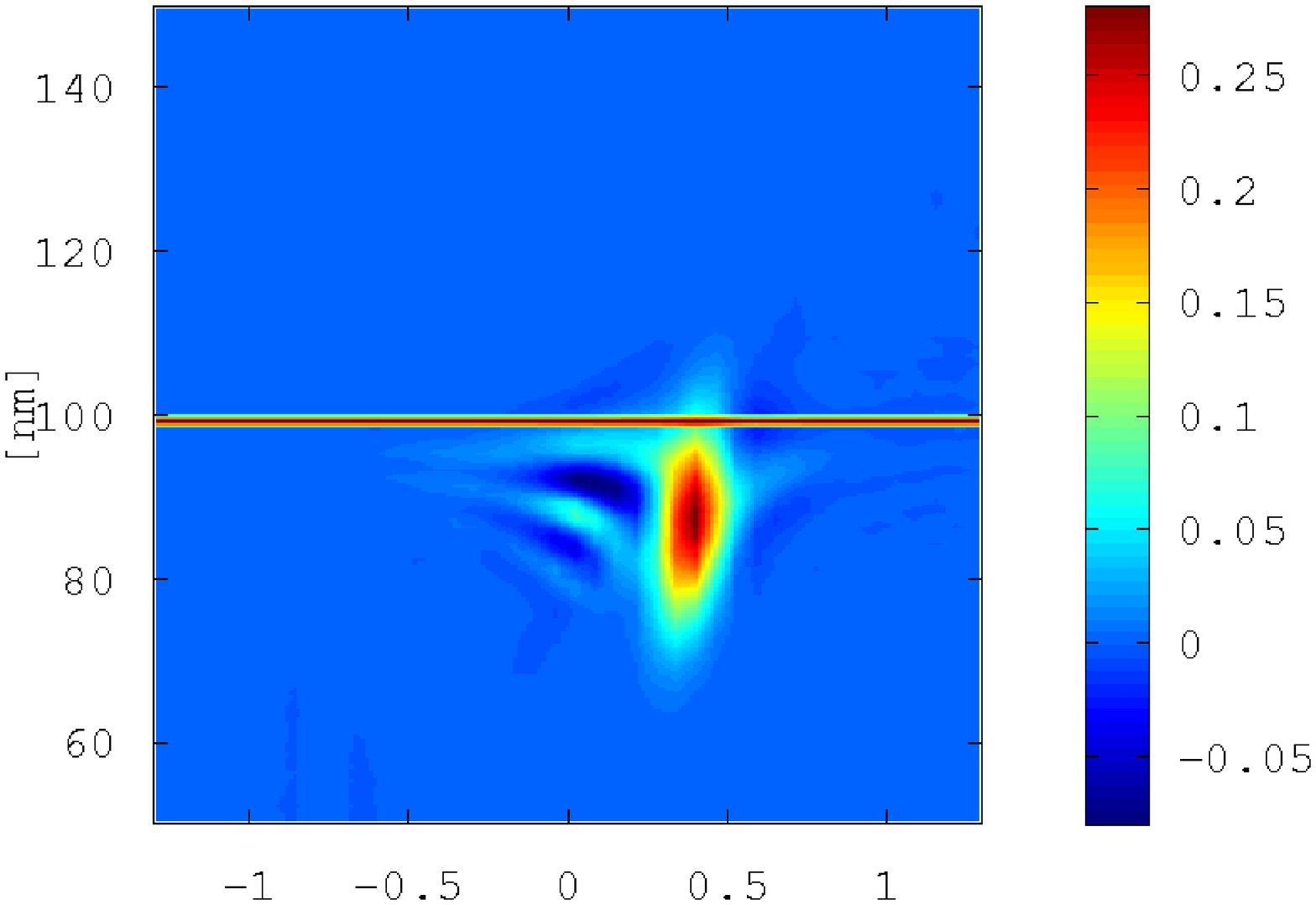}
\\
\includegraphics[width=0.45\textwidth]{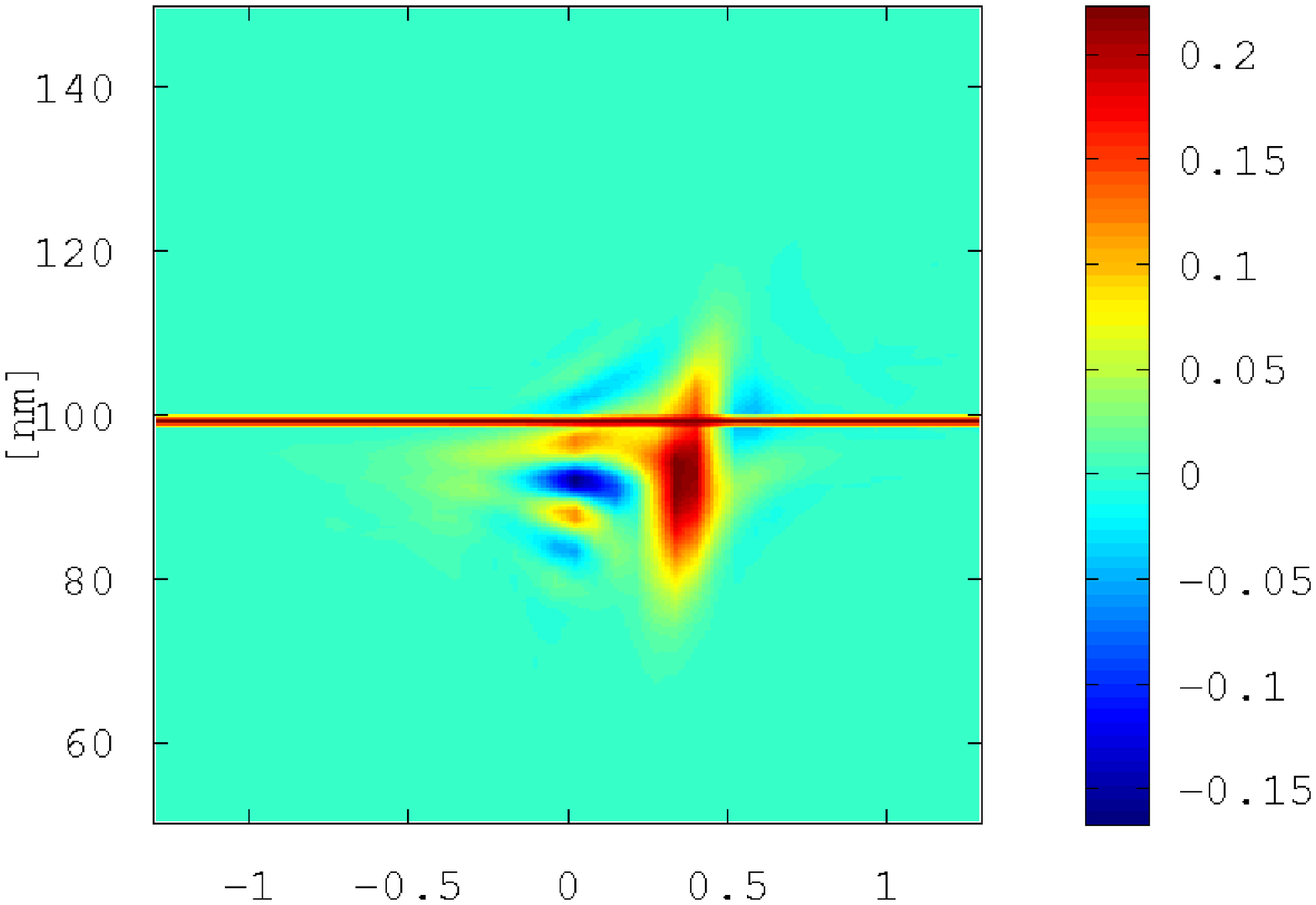}
\includegraphics[width=0.45\textwidth]{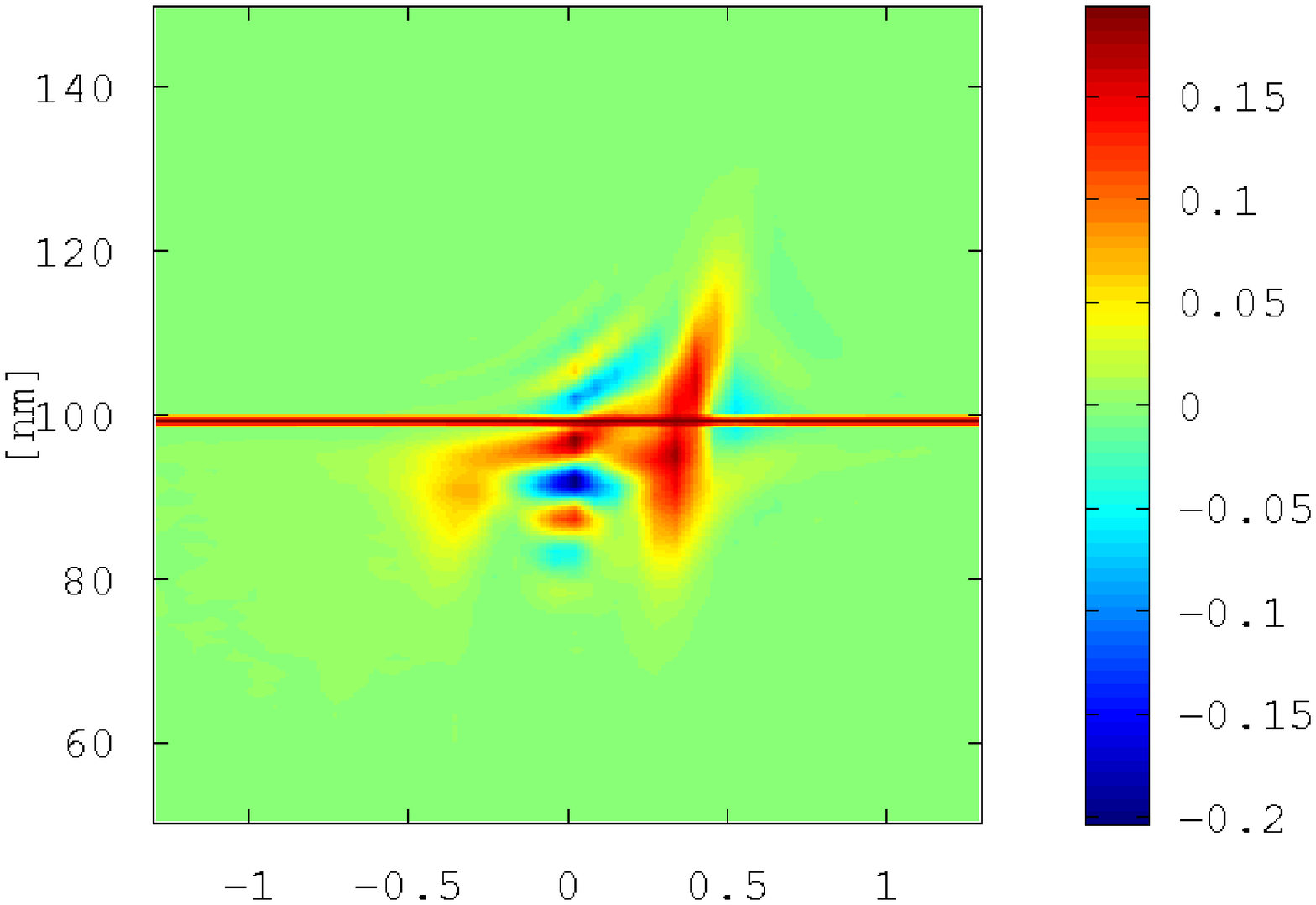}
\\
\includegraphics[width=0.45\textwidth]{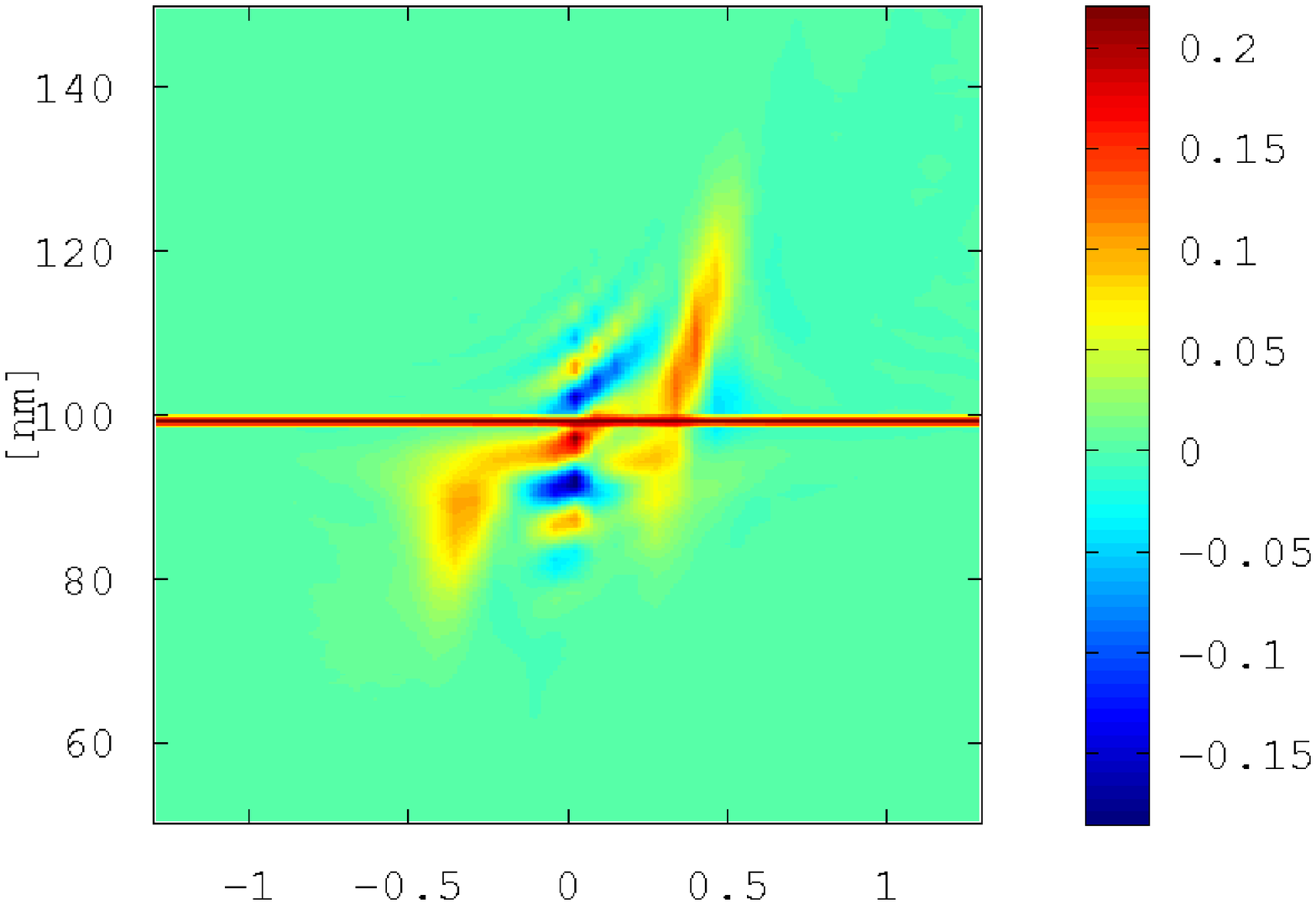}
\includegraphics[width=0.45\textwidth]{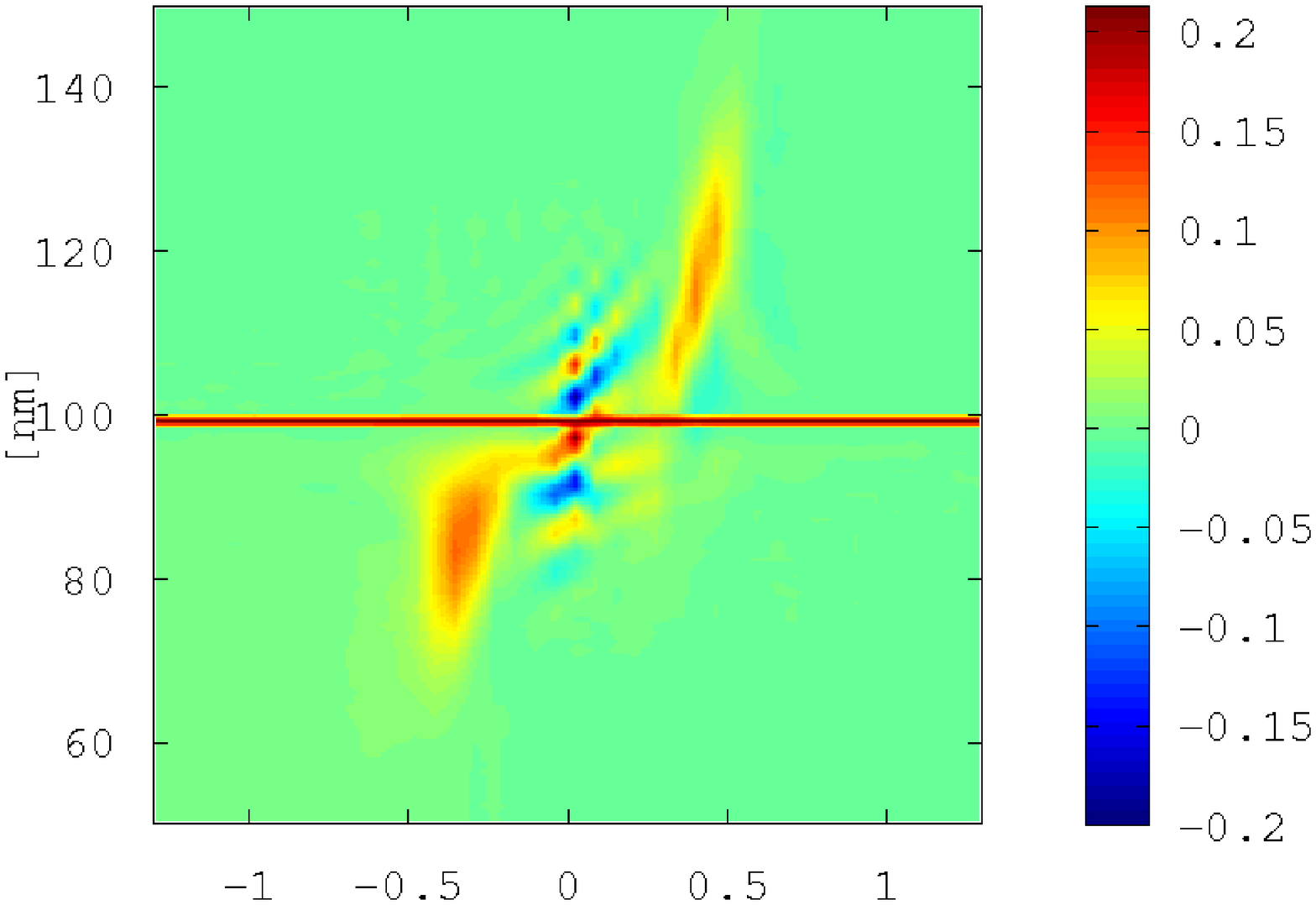}
\end{tabular}
\end{minipage}
\caption{Quantum tunnelling experiment in phase-space (from \cite{SPF}): an initially Gaussian wave-packet is directed towards a potential barrier equal to $0.10$ eV. The wave packet is represented by a set of signed particles (initially all positive) which is evolved in time. The times reported are (from left to right, from top to bottom) $0$ fs, $10$ fs, $20$ fs, $30$ fs, $40$ fs, $50$ fs, $60$ fs and $70$fs respectively. Negative values (dark blue) appear during the evolution. The position of the potential barrier is schematically represented by the (red) line in the middle of the domain.}
\label{distribution_01}
\end{figure}

Finally, the reader should note that the integral in (\ref{momentum_integral}) is not a Riemann integral.
In fact, we call it a {\sl{momentum integral}} by analogy with the concept of path integral (the former
covering the space of {\sl{momenta}}, the latter spanning the space of {\sl{paths}}). Results coming
from a practical implementation of the approach are shown in Fig. \ref{distribution_01} for an abrupt potential barrier.

\bigskip

{\sl{Classical limit, $\hbar \to 0$}}. In terms of the classical limit, there are some particular facets worth discussing.
It is indicative to note that the creation of signed particles is not allowed in the classical limit \cite{SPF}.
As such, we differentiate between two possible scenarios - the free space and systems in the presence of an external potential.
In the first case, the classical limit is trivial since the function $\gamma(x)$ is zero everywhere and, thereupon, no
particle creation event transpires at all. In the second scenario, however, i.e. when the potential $V(x) \neq 0$ for some $x$,
by introducing the change of integration variable $x' = \frac{p}{2 m} t'$ and supposing that the potential $V=V(x)$ can be
written in terms of a McLaurin series, we obtain the following expression
for the Wigner kernel (\ref{wigner-kernel})
\begin{eqnarray}
 V_W(x; p) &=& \frac{2 p}{\pi \hbar^2 m} \sum_{n=0, even}^{+\infty} \int_{-\infty}^{+\infty} dt' \cos(\frac{p^2 t'}{\hbar m})
\frac{V^{(n)}(x)}{n!} \left( \frac{p t'}{2m} \right)^n \nonumber \\
&=& \frac{2 p}{\pi \hbar^2 m} \int_{-\infty}^{+\infty} dt' \cos(\frac{p^2 t'}{\hbar m}) V(x) + \nonumber \\
&+& \frac{2 p}{\pi \hbar^2 m} \int_{-\infty}^{+\infty} dt' \cos(\frac{p^2 t'}{\hbar m}) \frac{1}{2} \frac{\partial^2 V(x)}{\partial x^2} \left( \frac{p t'}{2m} \right)^2 + \nonumber \\
&+& O(\hbar^4), \nonumber
\end{eqnarray}
keeping in mind that the summation is performed over even numbers, and $V^{(n)}(x)$ represents the
$n-$th order derivative of $V(x)$. Moreover, by omitting every term down to the second order with
respect to $\hbar$ and integrating, one acquires 
\begin{eqnarray}
V_W(x; p) &\approx& \left[ \frac{2 V(x)}{\pi p \hbar} \sin(\frac{p^2 t'}{\hbar m}) \right]_{-\infty}^{+\infty} \nonumber \\
          &=& 2 \lim_{t' \rightarrow +\infty} \frac{2 V(x)}{\pi p \hbar} \sin(\frac{p^2 t'}{\hbar m}) \nonumber \\
          &\approx& \frac{4 V(x)}{\pi p \hbar} \sin(\frac{p^2 T}{\hbar m}), \nonumber
\end{eqnarray}
for a {\sl{"long enough time"}} $T$.

Finally, through the combination of this result with formula (\ref{momentum_integral}) we obtain
the corresponding function $\gamma(x)$ which, now, reads:
$$
 \gamma(x) = \lim_{\Delta p \rightarrow 0^+} \sum_{M = -\infty}^{+\infty} \left[ \frac{4 V(x)}{\pi \hbar (M \Delta p)}
\sin(\frac{(M \Delta p)^2 T}{\hbar m})  \right]^+,
$$
with $[ \dots ]^+$ depicting the positive part of the quantity in the brackets.

For the sake of simplicity, supposing that $V(x)>0$ for $\forall x$, the term in brackets is transformed to positive
only when
$$
 0 \le M \le \frac{\hbar m \pi}{\Delta p^2 T},
$$
and, for $\hbar \rightarrow 0$, one has $$ 0 \le M \le 0 $$ hence $M=0$. 
In other words, $\gamma(x)=0$ for $\forall x$, with creation of particles forbidden. The reader should note that the
generalization of this result to the many-body case is obtained in a similar manner and leads to the same conclusion.

\bigskip

\begin{figure}[h!]
\centering
\begin{minipage}{1.0\textwidth}
\centering
\begin{tabular}{c}
\includegraphics[width=0.99\textwidth]{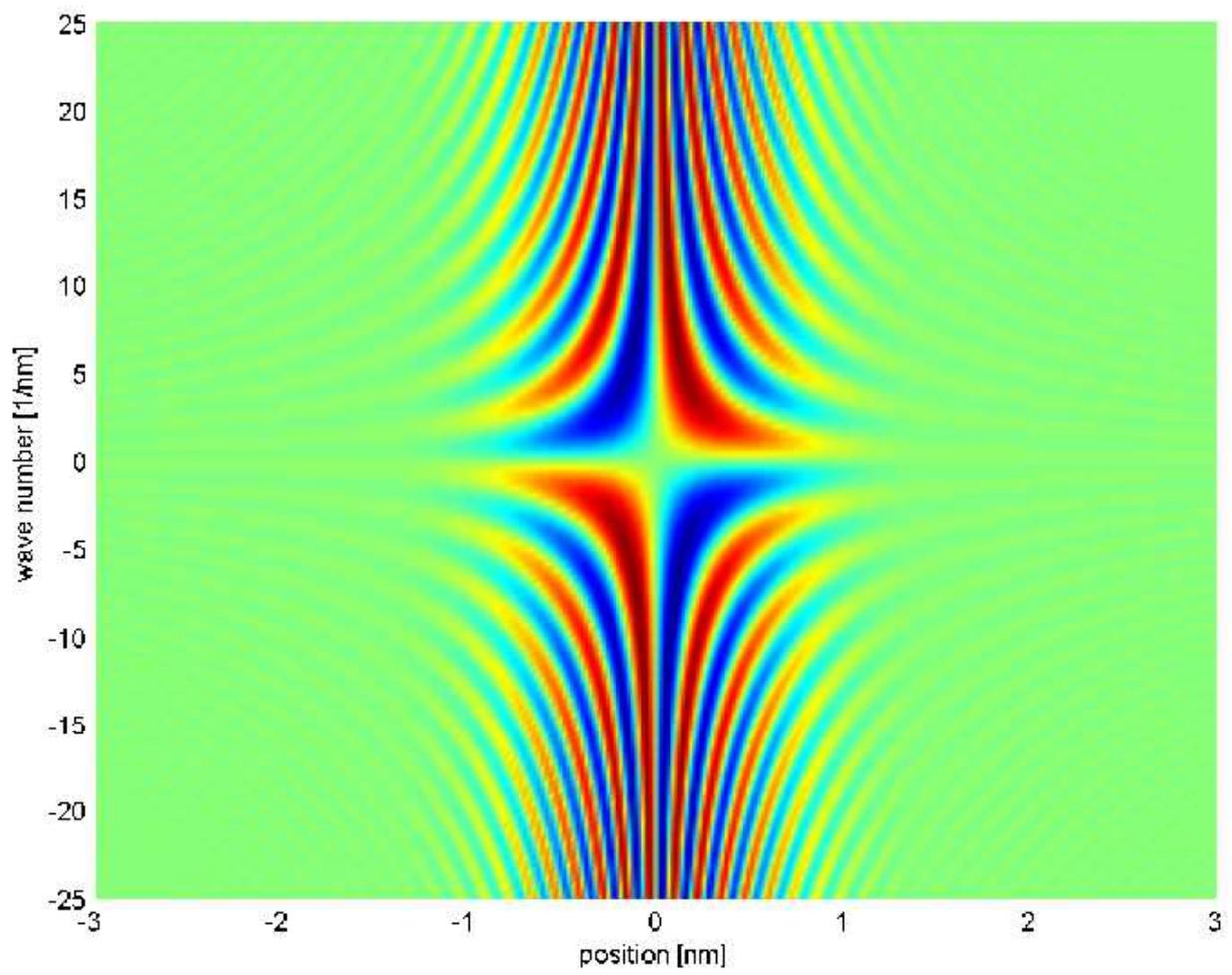}
\end{tabular}
\end{minipage}
\caption{The Wigner kernel $V_W(x;p)$ corresponding to a Gaussian potential barrier.}
\label{wigner_kernel}
\end{figure}

{\sl{Quantum Tunnelling}}. The concept of signed particle formulation with its three postulates provides a new physical picture, distinctive
and peculiar compared to the other available formulations of quantum mechanics. Important example to consider comes
from the phenomenon of quantum tunnelling. In this work, we consider a one-dimensional electron impinging on a Gaussian
potential barrier centered at the origin $x=0$:
\begin{equation}
 V(x) = -U e^{-\frac{x^2}{2 \sigma^2}},
\end{equation}
with $U$ being a negative value and representing the peak of the barrier, and $\sigma$ expressing the dispersion of
the barrier in the real space. In this simple scenario, one can demonstrate by means of standard analytical
manipulations of formula (\ref{wigner-kernel}), that the Wigner kernel reads:
\begin{equation}
 V_W(x;p) = +4 \sqrt{2 \pi} U \sigma e^{-2 \left( \frac{\sigma p}{\hbar} \right)^2} \sin{\left(\frac{2 x p}{\hbar} \right)},
\end{equation}
(a representation is reported in Fig. \ref{wigner_kernel}).
Actualizing the fact that $ \left[ f(x) \right]^+ = \frac{\bigl|f(x)\bigr|+f(x)}{2}$, one can show that:
\begin{eqnarray}
 \left[ V_W(x; p) \right]^+ &=& -4 \sqrt{2 \pi} U \sigma e^{-2 \frac{\sigma^2 p^2}{\hbar^2}} \times \left[ \sin{\left(\frac{2 x p}{\hbar} \right)} \right]^+ \nonumber \\
 &=& -4 \sqrt{2 \pi} U \sigma e^{-\frac{2 p^2 \sigma^2}{\hbar^2}} \left\{ \frac{\bigl| \sin{\left(\frac{2 x p}{\hbar} \right)} \bigr| + \sin{\left(\frac{2 x p}{\hbar} \right)}}{2}  \right\},
\end{eqnarray}
which, by utilizing (\ref{momentum_integral}) and the antisymmetry of the sinusoidal function, finally leads to: 
\begin{equation}
 \gamma(x) = -4 \sqrt{2 \pi} U \sigma \times \lim_{\epsilon \to 0} \sum_{M \in \mathbb{N}} \left\{ e^{-2 \frac{\sigma^2 (M \epsilon)^2}{\hbar^2}} \bigl| \sin{\left(\frac{2 x M \epsilon}{\hbar} \right)} \bigr|\right\}.
\label{series}
\end{equation}

It is important to note that the above series can be shown to be convergent by using the following series of inequalities. Indeed, one gets:
$$
 0 < e^{-\frac{2 M^2 \epsilon^2 \sigma^2}{\hbar^2}} \bigl| \sin{\left(\frac{2 x M \epsilon}{\hbar} \right)}  \bigr| < e^{-\frac{2 M^2 \epsilon^2 \sigma^2}{\hbar^2}} < \frac{1}{\left( 2 M^2 \epsilon^2 \sigma^2 \right)!}.
$$

\bigskip

\begin{figure}[h!]
\centering
\begin{minipage}{1.0\textwidth}
\centering
\begin{tabular}{c}
\includegraphics[width=0.5\textwidth]{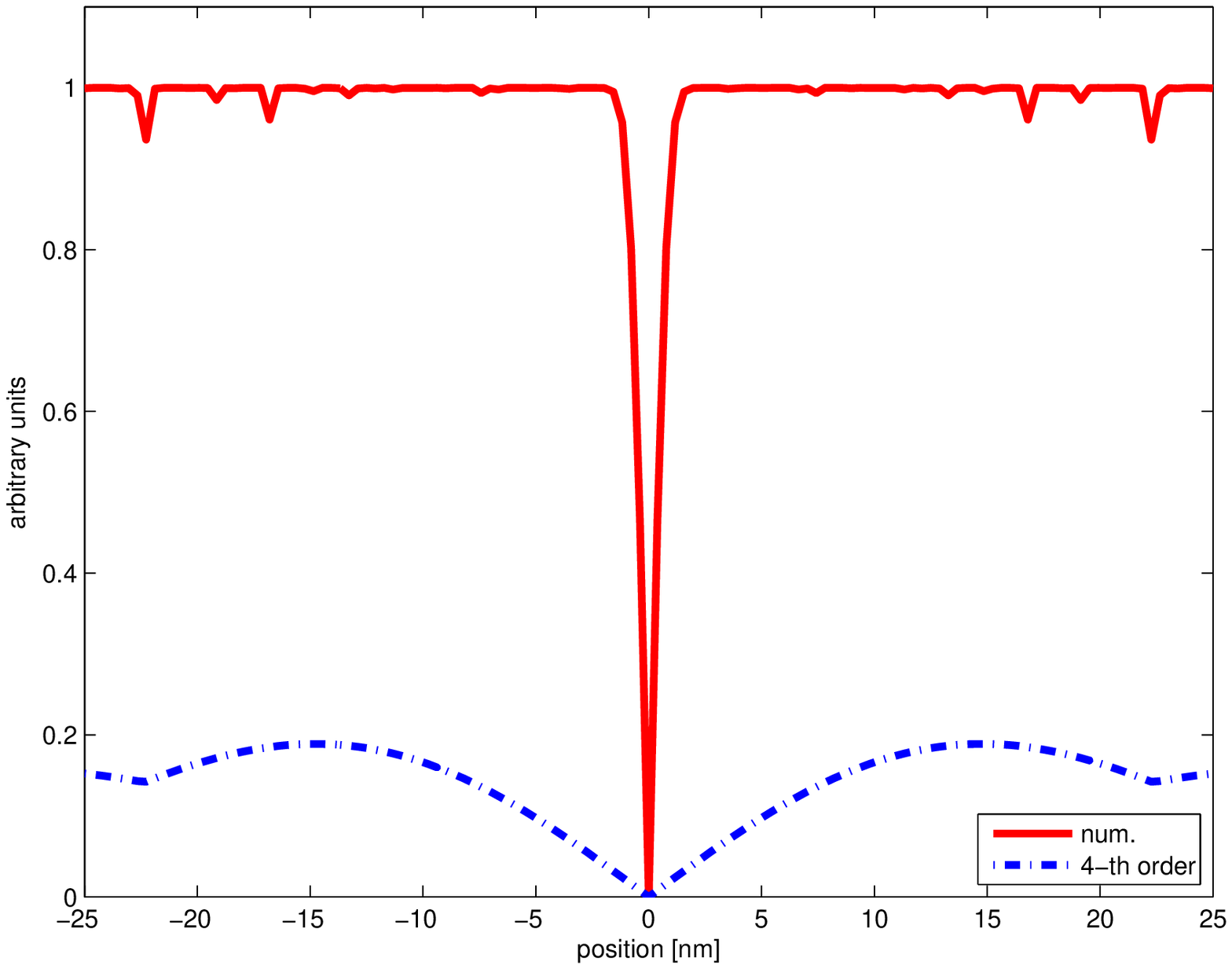}
\includegraphics[width=0.5\textwidth]{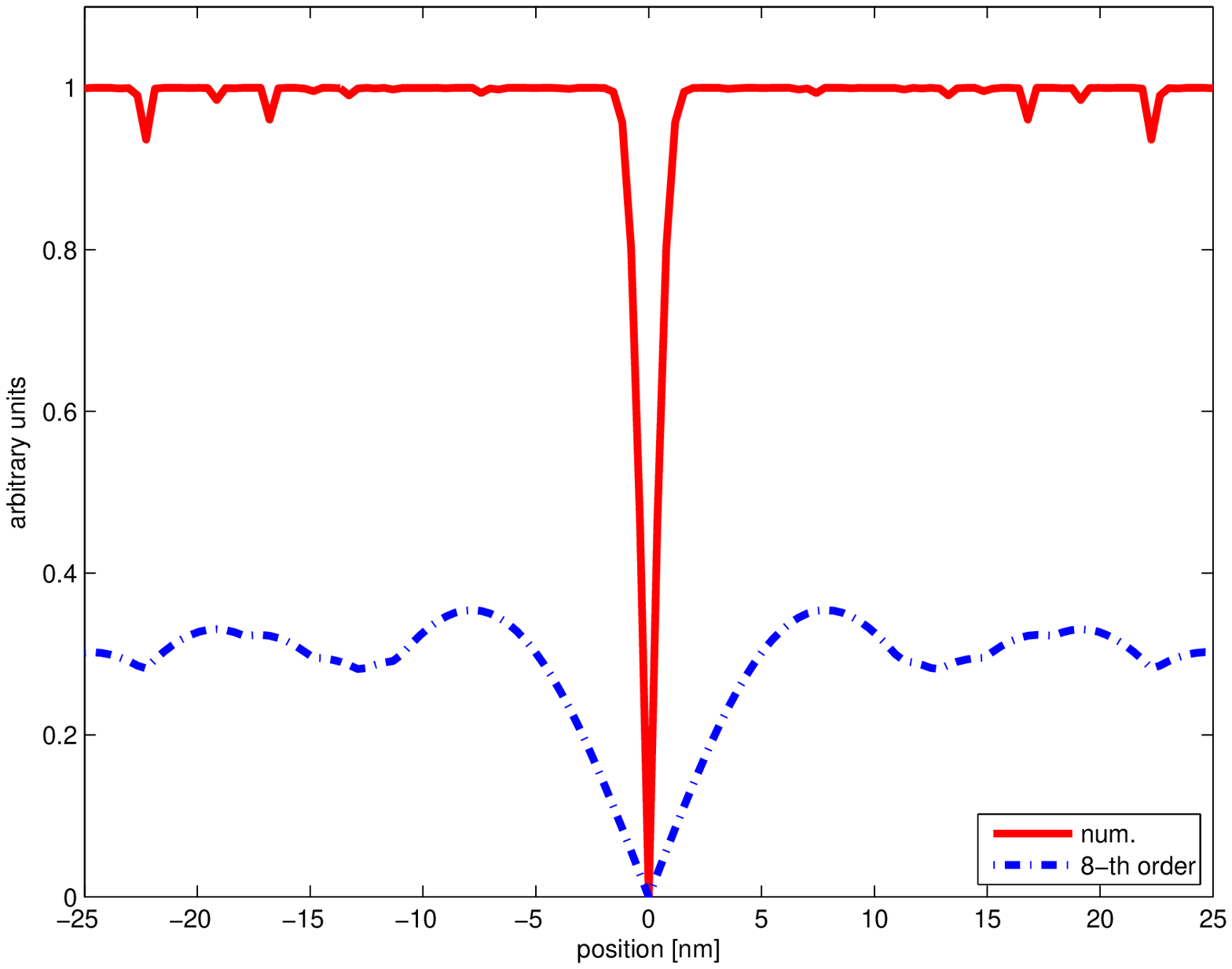}
\\
\includegraphics[width=0.5\textwidth]{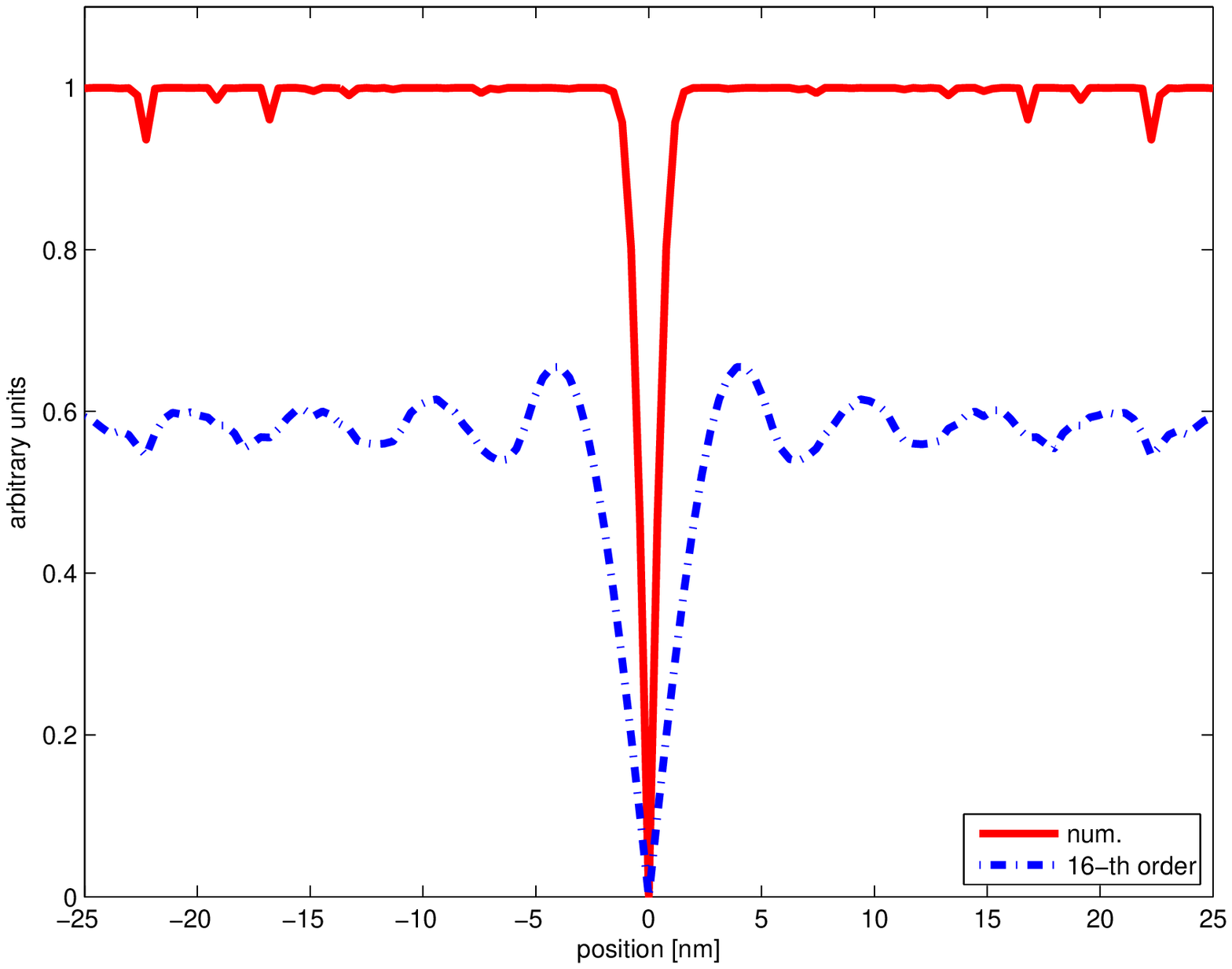}
\includegraphics[width=0.5\textwidth]{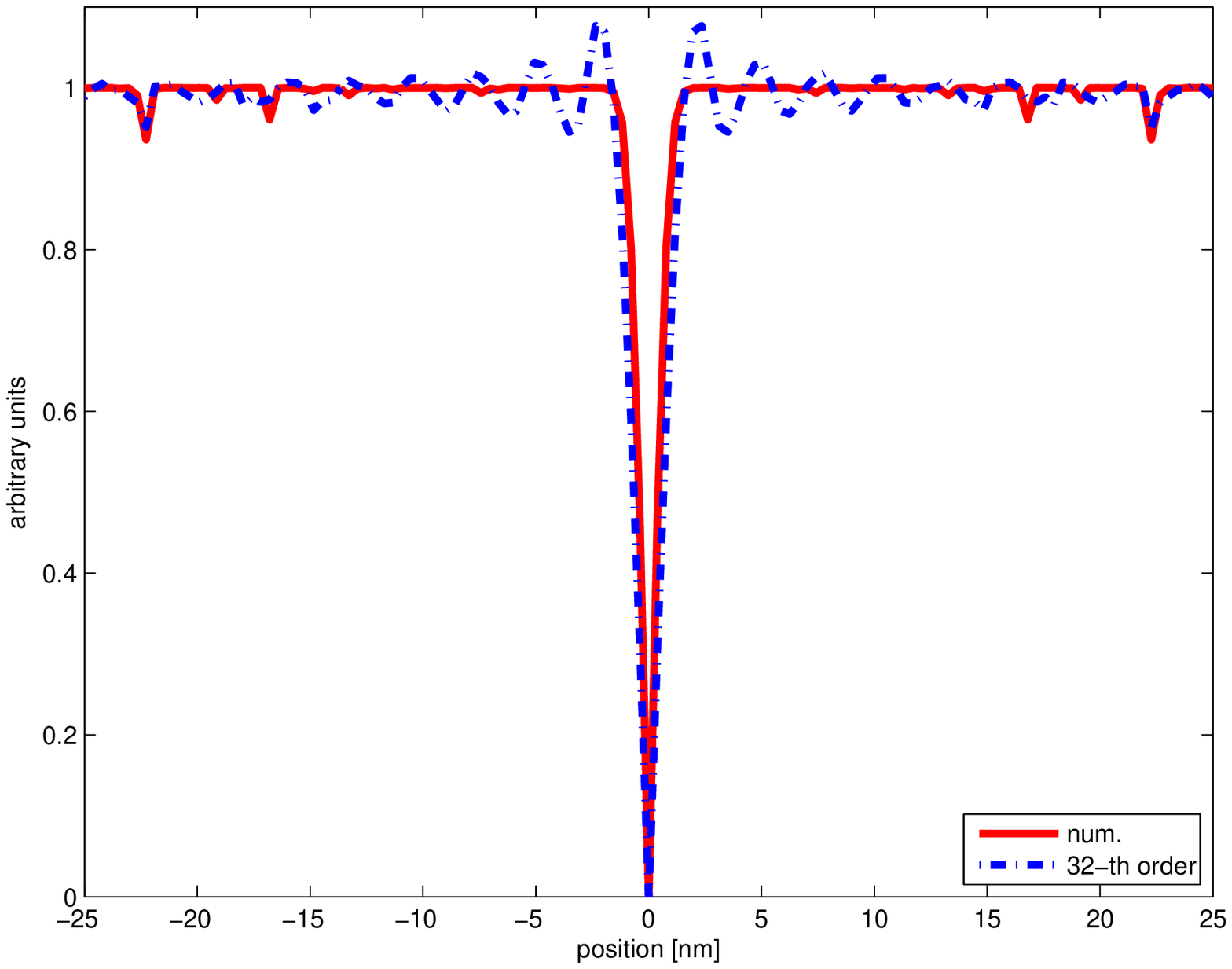}
\end{tabular}
\end{minipage}
\caption{The function $\gamma(x)$ in the case of a Gaussian potential barrier.}
\label{gamma_functions}
\end{figure}

Additionally, to enable a physical interpretation of the function $\gamma=\gamma(x)$, we now explicit a few terms of the series (\ref{series}).
In turn it provides:
\begin{eqnarray}
 \gamma(x) = -4 \sqrt{2 \pi} U \sigma \times \lim_{\epsilon \to 0} && \left\{ e^{-2 \frac{\sigma^2 (\epsilon)^2}{\hbar^2}} \bigl| \sin{\left(\frac{2 x \epsilon}{\hbar} \right)} \bigr| + e^{-2 \frac{\sigma^2 (2 \epsilon)^2}{\hbar^2}} \bigl| \sin{\left(\frac{4 x \epsilon}{\hbar} \right)} \bigr| \right. \nonumber \\
&+& \left. e^{-2 \frac{\sigma^2 (3 \epsilon)^2}{\hbar^2}} \bigl| \sin{\left(\frac{6 x \epsilon}{\hbar} \right)} \bigr| + e^{-2 \frac{\sigma^2 (4 \epsilon)^2}{\hbar^2}} \bigl| \sin{\left(\frac{8 x \epsilon}{\hbar} \right)} \bigr| \right. \nonumber \\
&+& \left. \dots \right\}.
\end{eqnarray}
It becomes immediately apparent that the exponentials in the formula above tend to damp the various terms with an increasing value $M$.
One should note that the damping is not rapid. In fact, for very small $\epsilon$ the two quantities $\sigma M \epsilon$ and $\hbar$ tend
to be comparable. As a consequence, the decrease of the series is actualized at a very slow rate, with many terms needing to be taken
in the series in order to have a good approximation of the function $\gamma=\gamma(x)$. On the other side, the
terms $\bigl| \sin{\left(\frac{2 x M \epsilon}{\hbar} \right)} \bigr|$ contribute with values in the range $[0,1]$ where 
one can see how these oscillating contributions sum up in Fig. \ref{gamma_functions}. The numerically computed function $\gamma(x)$
is depicted by a (red) continuous curve, which is compared  to various sums of terms up to $4$, $8$, $16$ and $32$ respectively,
represented by the (blue) dashed curves. It is clear that despite the slow convergence, the series eventually get closer and
closer to the numerical $\gamma(x)$ function. One important observation in regards to the quite unexpected spikes close to
the boundaries of the observable domain are actually approached by the converging series of terms. Although there exist a
possibility for their incorrect interpretation as numerical spurious oscillations.

Additionally, it is feasible to extract several mathematical properties for the function $\gamma(x)$ which would hardly be visible in a purely
numerical context. Indeed, one can easily show that $\gamma(x)=\gamma(-x)$ for all $x \in ]-\infty, +\infty[$ and that $\gamma(0)=0$.
We should point out that these properties have an interesting physical interpretation. The first one defines the system
as symmetric with respect to the creation of signed particles, and something rather presupposed from a symmetric potential
(the Gaussian barrier is symmetric with respect to $x=0$). The more interesting case comes with the property $\gamma(0)=0$,
which elucidates a physical fact otherwise impossible to distinguish through purely numerical arguments. One
observes in this case that particle creation is forbidden at the top of the Gaussian barrier and signed particles
going through that (very restricted but apparently non-negligible) area behave as field-less classical objects. 
As such, the area around the top behaves just like a (short) domain with a constant applied potential.

As discussed in the examples above, the Signed particle formulation provides a possible level of detail never
attained before. We are now ready to apply the formulation to the simulation of the hydrogen atom as a full
quantum system, as discussed in the next section.

\section{Beyond the Born-Oppenheimer approximation}

\begin{figure}[h!]
\centering
\begin{minipage}{1.0\textwidth}
\begin{tabular}{c}
\includegraphics[width=0.5\textwidth]{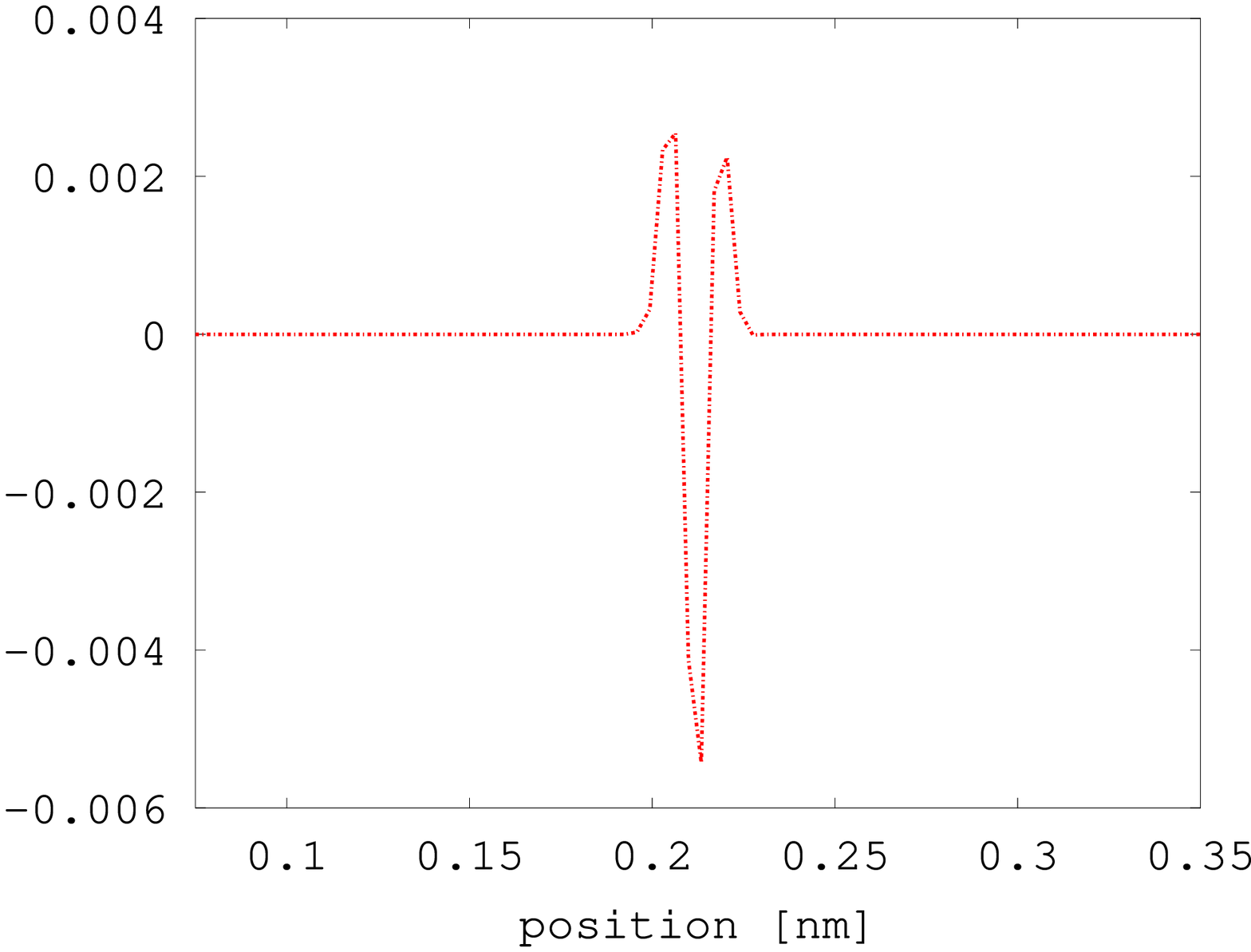}
\includegraphics[width=0.55\textwidth]{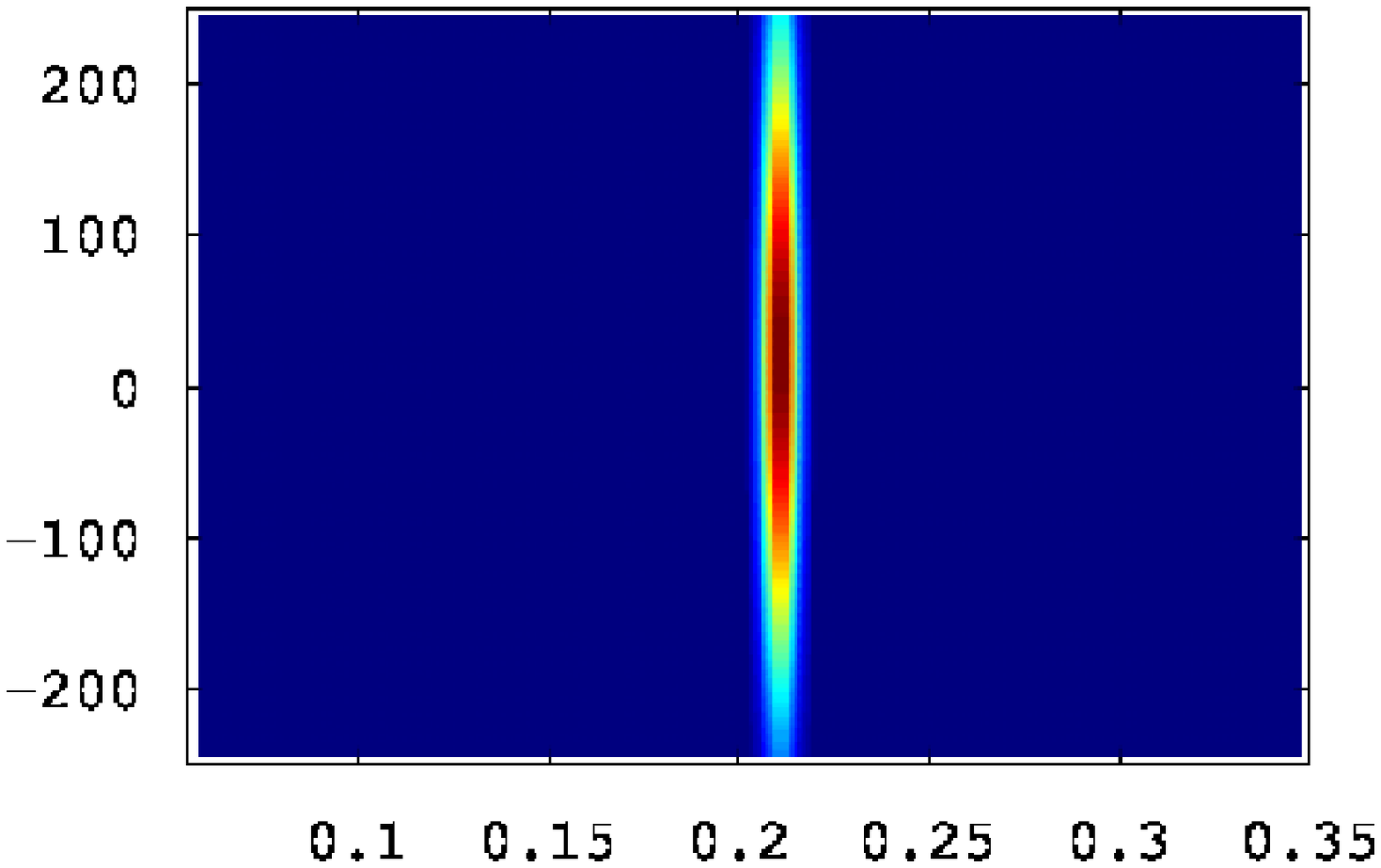}
\end{tabular}
\end{minipage}
\caption{Left: Relative difference corresponding to the proton probability densities $\rho_0(x)$ and $\rho(x;t)$ at times, respectively, $t=0$ as and $t=6$ as. The difference is defined as $\varepsilon (x) = \frac{\rho(x; t=6) - \rho_0(x)}{\max_{x \in D} \rho_0(x)}$, with $D$ the spatial domain of the simulation. Right: initial distribution function for the proton. The positions (x-axis) are expressed in nanometer while the wave-numbers (y-axis) are in nanometer$^{-1}$.}
\label{difference}
\end{figure}

\begin{figure}[h!]
\centering
\begin{minipage}{1.0\textwidth}
\begin{tabular}{c}
\includegraphics[width=0.5\textwidth]{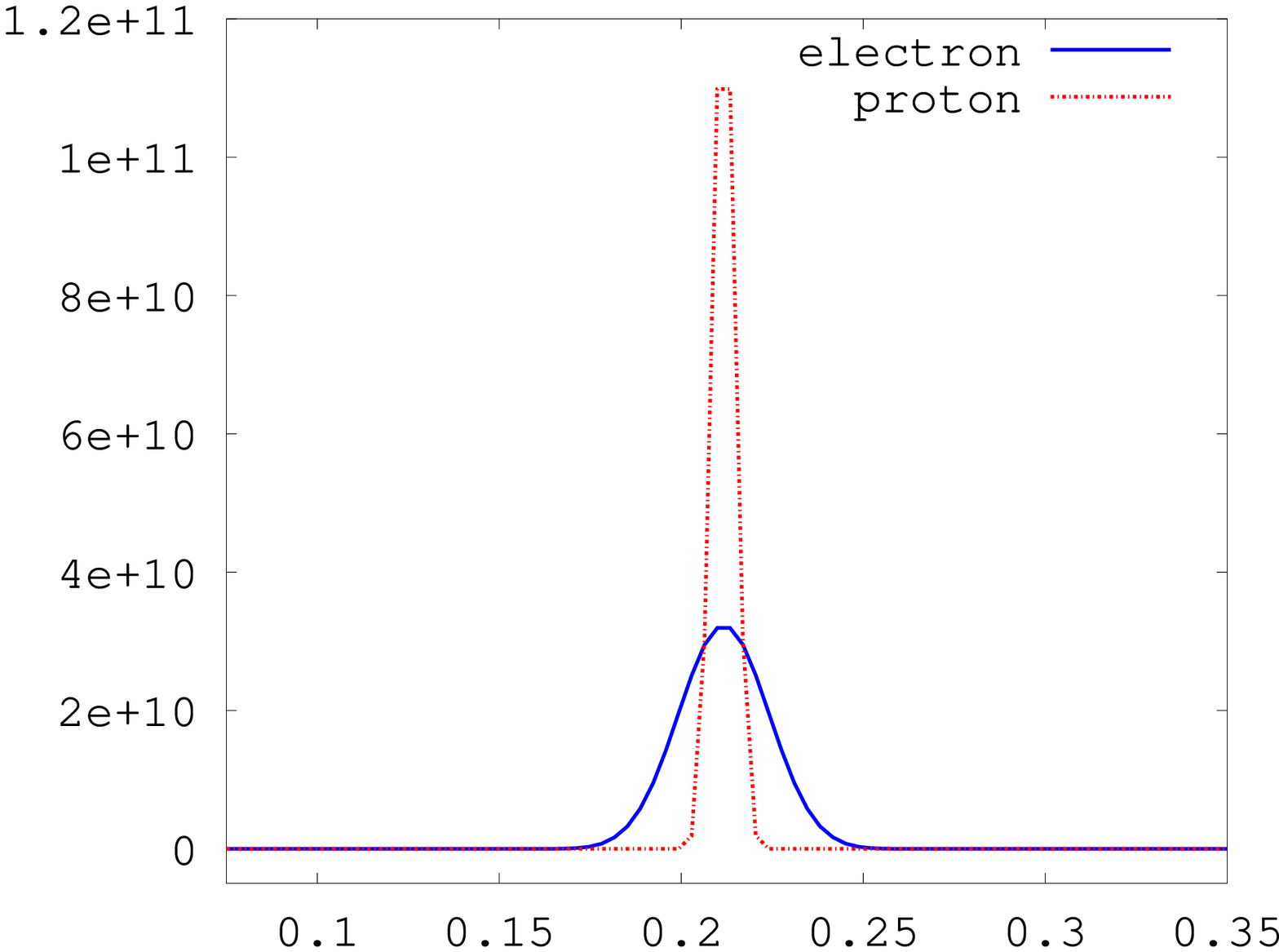}
\includegraphics[width=0.55\textwidth]{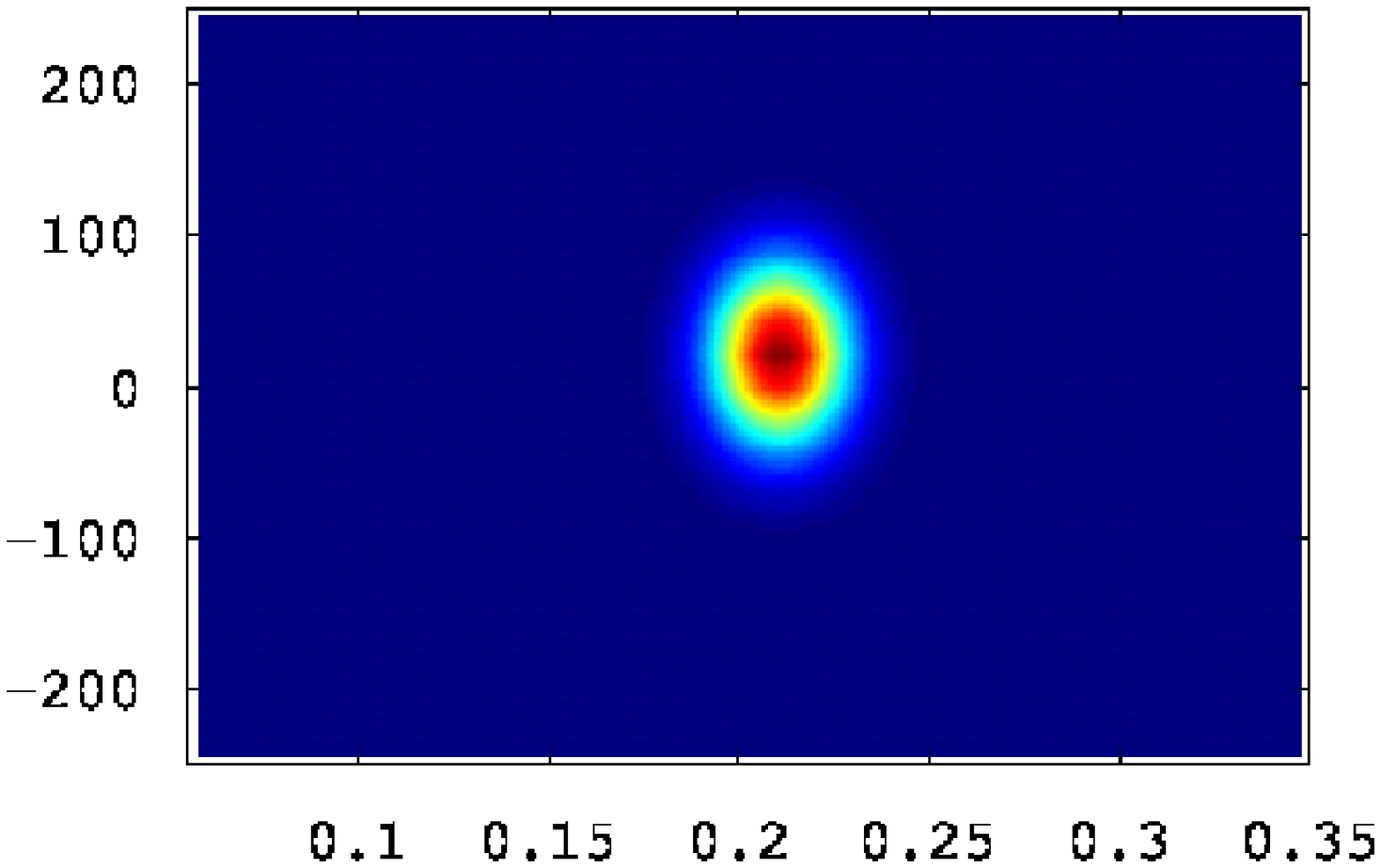}
\\
\includegraphics[width=0.5\textwidth]{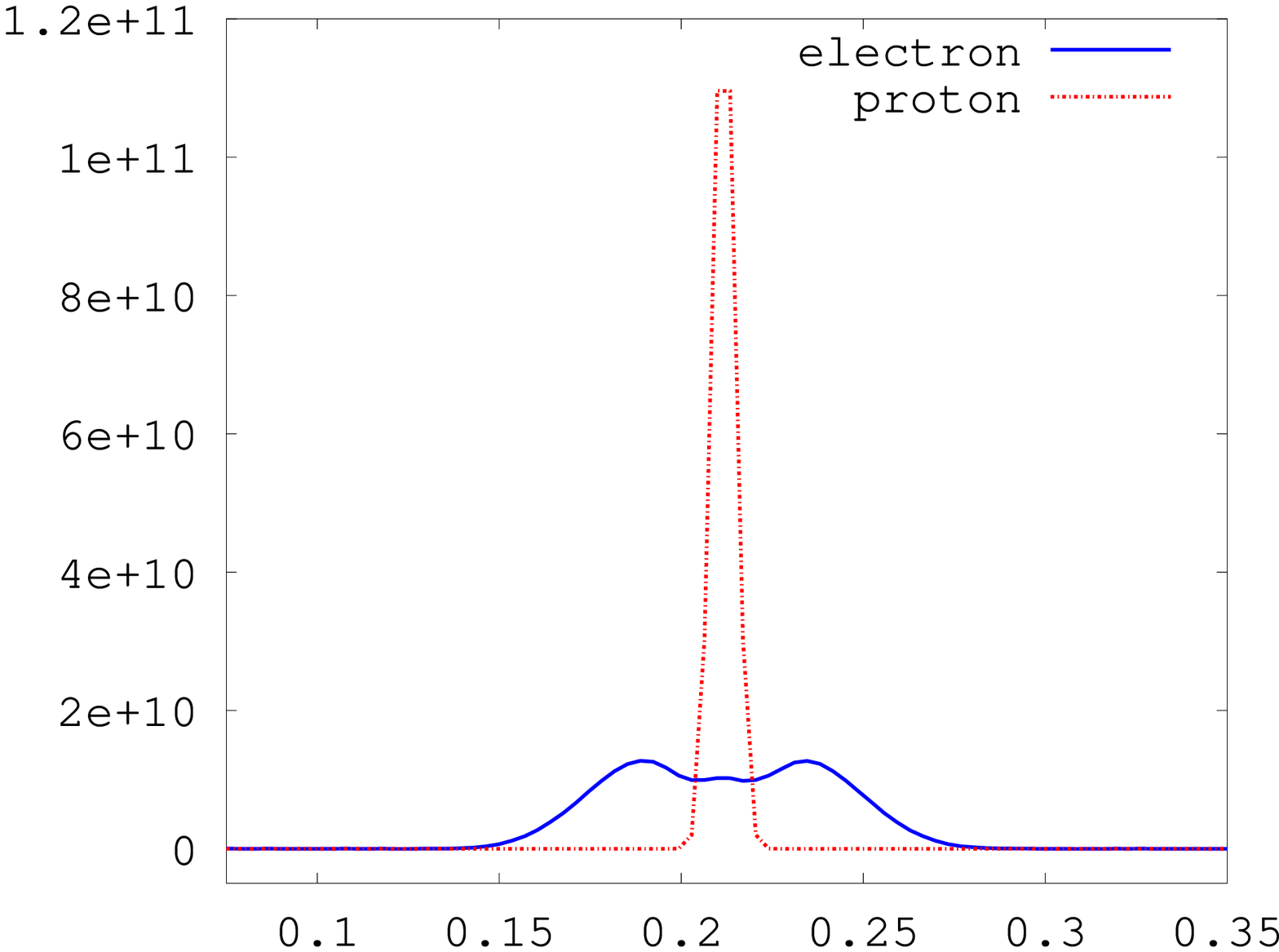}
\includegraphics[width=0.55\textwidth]{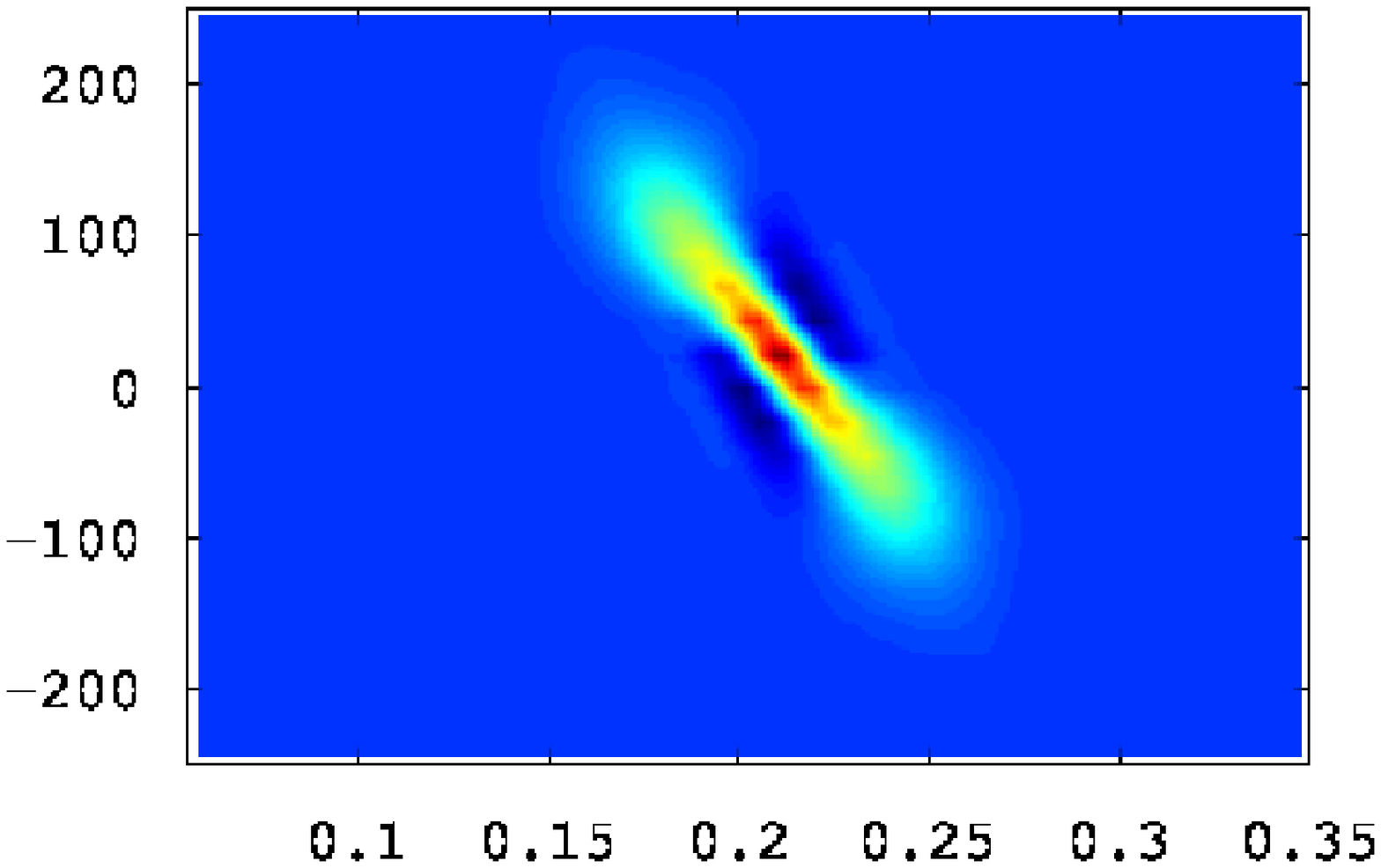}
\\
\includegraphics[width=0.5\textwidth]{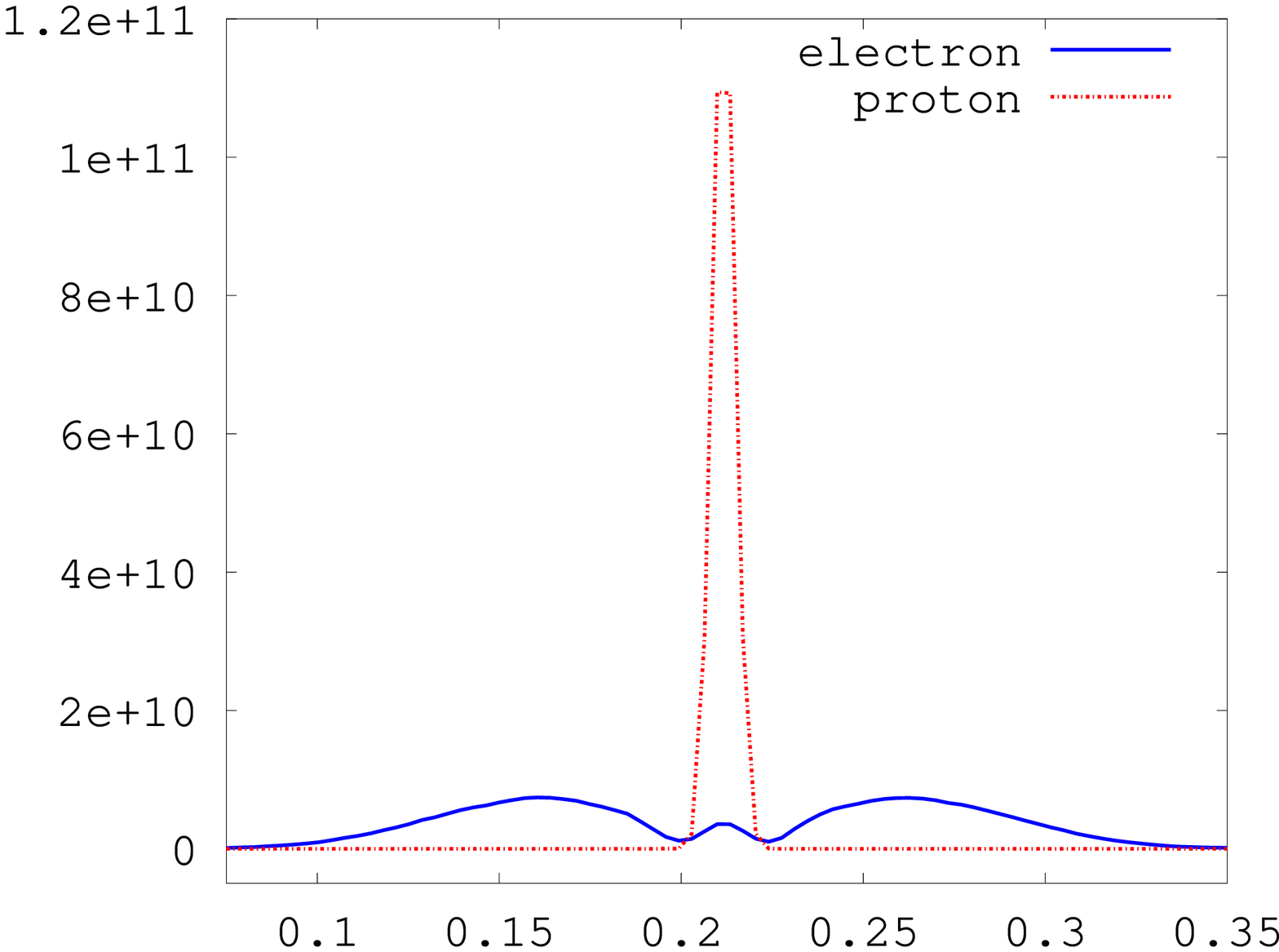}
\includegraphics[width=0.55\textwidth]{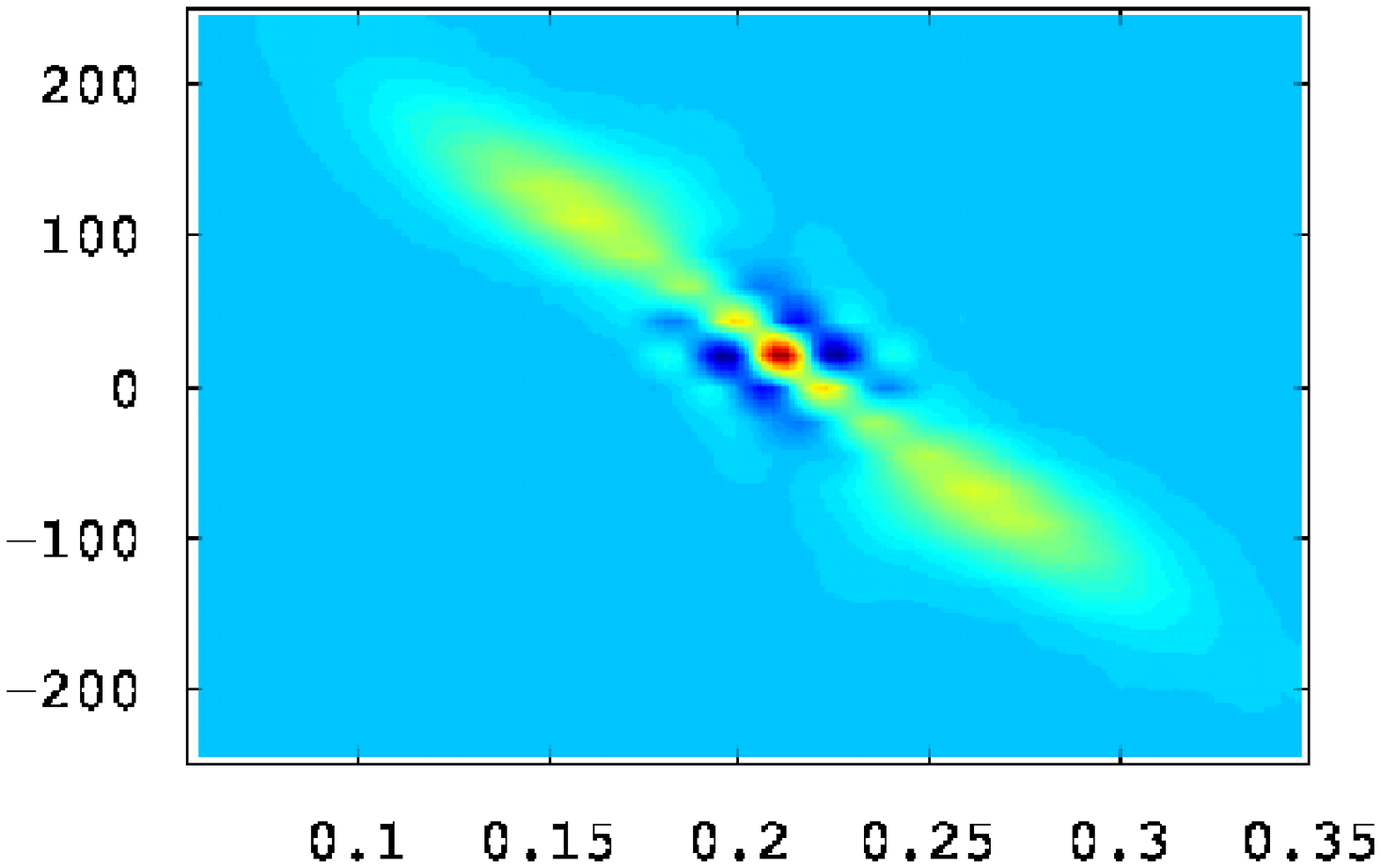}
\end{tabular}
\end{minipage}
\caption{Left column: time-dependent evolution of the (normalized) probability density for the electron, (blue) continuous curve, and proton, (red) dashed curve. The curves (from top to bottom) correspond to the times $0$ as, $3$ as and $6$ as respectively. The appearance of a radius comparable to the Bohr radius is observable after a few attoseconds in the probability density of the electron. Positions (x-axis) are in nanometers. Right column: evolution in time of the electron quasi-distribution function. Negative peaks appear after several attoseconds, a clear signature of the presence of the Heisenberg principle. Positions (x-axis) are in nanometers, Wave-numbers (y-axis) are in $1/$nanometer.}
\label{probability-density}
\end{figure}

Considering the hydrogen atom as a (inseparable) two-body system consisting of a (quantum) proton and a (quantum) electron,
its time-dependent dynamics is governed by the two-body Wigner equation (in one-dimensional space and Bloch formalism):
\begin{equation}
 \frac{\partial f_W}{\partial t} + \frac{\hbar k_{e^-}}{m_{e^-}} \frac{\partial f_W}{\partial x_{e^-}} + \frac{\hbar k_{p^+}}{m_{p^+}} \frac{\partial f_W}{\partial x_{p^+}} = \hbar^2 \int d{\bf{k}}' f_W({\bf{x}}; {\bf{k}}') V_W({\bf{x}}; {\bf{k}} + {\bf{k}}'),
\end{equation}
a partial integro-differential equation where $x_{e^-}$ and $x_{p^+}$ are the position of the electron and proton respectively,
with ${\bf{x}} = (x_{e^-}, x_{p^+})$, $k_{e^-}$ and $k_{p^+}$ representing the wave-number of the electron and proton respectively,
with ${\bf{k}} = ( k_{e^-}, k_{p^+})$, $f_W = f_W({\bf{x}}; {\bf{k}})$ the Wigner quasi-distribution function of the system,
and $V_W=V_W({\bf{x}}; {\bf{k}})$ the Wigner kernel expressing the forces involved along with their non-local nature \cite{Wigner}.
If we consider the results from the previous section, we can define the initial conditions for both particles to consist of 
Gaussian wave-packets where the proton is highly localized in space (and therefore strongly delocalized in wave-space), 
see Fig. \ref{difference} (right-hand side), and the electron is delocalized in both space and wave-space
(around $k=0/$nm), see Fig. \ref{probability-density} (top, right-hand side).

The reader can see the results of the simulation, presented in Figs. \ref{probability-density} and \ref{difference}.
The left column of Fig. \ref{probability-density} shows the time-dependent evolution of the normalized
probability density for the electron (with blue- continuous curve), and for the proton (with red- dashed curve).
The times reported (from top to bottom) correspond to $0$, $3$ and $6$ attoseconds respectively.
Interestingly, one can notice the emergence of a radius for the electron probability density
comparable to the Bohr radius ($\approx 0.529$ Angstrom) after a few attoseconds. 
The evolution in time of the very same electron in the phase-space (in other words, the electron
quasi-distribution function) is shown in the right column of Fig. \ref{probability-density}.

The observed negative peaks appearing near the point \hfill \break 
$(x,k) = (0.20991 nm, -22.263/nm)$, represent
a definite indication of the Heisenberg principle. While the electron dynamics is very fast 
(on a time scale of the order of attoseconds), Fig. \ref{difference} indicates how the evolution
of the proton proceeds comparatively very slowly. Distinctly, the left-hand side plot demonstrates
the relative difference between the proton probability densities $\rho_0(x)$ and $\rho(x;t)$ at
times $t=0$ as and $t=6$ as, respectively. Here the difference is defined as:
\begin{equation}
 \varepsilon (x_{p^+}) = \frac{\rho(x_{p^+}; t=6) - \rho_0(x_{p^+})}{\max_{x_{p^+} \in D} \rho_0(x_{p^+})},
\end{equation}
with $D$ being the spatial domain of the simulation. While the plot
shows how slow the proton dynamics is, it, in fact, also appears that 
the proton wave-packet is evolving in time. This is conceivable due to the
vibrational interactions with the electron, an effect that apparently seems to
happen even in the hydrogen atom although it remains almost imperceptible.

\clearpage

\section{Conclusions}

In this paper, we discussed the application of the recently suggested formulation of quantum mechanics involving
the concept of signed particles to the time-dependent simulation of an hydrogen atom beyond
the Born-Oppenheimer approximation. We have provided an example studying quantum tunnelling trough a simple Gaussian barrier
in terms of the signed particle formulation. We have furthermore shown through the utilization of a set of
three simple postulates, how the hydrogen atom can be considered as a whole quantum two-body problem.
While this direct simulation is limited to the one-dimensional case, we can certainly draw interesting
conclusions about the results of the simulation. We expect that the experiments shown in this work
should bring insights into the field of non-adiabatic quantum simulations which are hardly reachable
with other methods, and should open a totally new perspective in the field of quantum chemistry. We
hope this will encourage other scientists to utilize this formalism for their own scientific investigations.

\bigskip

{\bf{Acknowledgements}}.
J.M.~Sellier acknowledges P. Dollfus, D. Querlioz and S. Shao for the inspiring discussions, and M. Anti for her support and encouragement. K.G.~Kapanova
has been partially supported by the "Research program for young scientists at the Bulgarian Academy of Science
2017", funded by the Bulgarian Ministry of Education and Science.

\bibliographystyle{elsarticle-num}
\bibliography{Sellier-JCP-2017.bib}

\end{document}